\documentclass[aps,prl,twocolumn,superscriptaddress,amsmath,amssymb,footinbib,showpacs]{revtex4-1}

\usepackage{bibunits}

\usepackage{pdfpages}

\newcommand{\Jnature}{Nature (London)}

\newcommand{\Jscience}{Science}

\newcommand{\Jprl}{Phys. Rev. Lett.}
\newcommand{\Jpr}{Phys. Rev.}
\newcommand{\Jpra}{Phys. Rev. A}
\newcommand{\Jprb}{Phys. Rev. B}

\newcommand{\Jpre}{Phys. Rev. E}

\newcommand{\Jrmp}{Rev. Mod. Phys.}

\newcommand{\Jepjb}{Eur. Phys. J. B}

\newcommand{\JphyslettA}{J. Phys. Lett. A}

\newcommand{\JJlowT}{J. Low Temp. Phys.}

\newcommand{\Jprocroysoc}{Proc. Roy. Soc. A: Math. Phys. Eng. Sci.}

\newcommand{\Jjetp}{Sov. Phys. JETP}
\newcommand{\Jjetplett}{JETP Lett.}

\newcommand{\JphysicaB}{Physica B}

\newcommand{\JjphysC}{J. Phys. C: Solid State Phys.}

\newcommand{\Jadvphys}{Adv. Phys.}

\usepackage[english]{babel}
\usepackage{latexsym}
\usepackage{graphics}
\usepackage{epsfig}
\usepackage{color}
\usepackage{hyperref}

\hypersetup{
colorlinks=true,
citecolor=blue,
linkcolor=red,
urlcolor=black
}
\usepackage{mathptmx}

\newcommand{\ie}{{i.e.}}

\newcommand{\kB}{k_\textrm{\tiny B}}

\def\dd{\,\mathrm{d}}

\def\Tr{\mathrm{Tr}}

\renewcommand{\vec}[1]{\mathbf{#1}}
\newcommand{\sub}[1]{\textrm{\tiny #1}}

\newcommand{\braket}[1]{\ensuremath{\langle \, #1 \, \rangle} \,}

\newcommand{\reffig}[1]{Fig.\,\ref{#1}}

\def\ns{n_\sub{s}}

\def\Er{E_\sub{r}}
\def\hopping{J}

\def\Kc{K_\sub{c}}

\def\ns{n_\sub{s}}
\def\fs{f_\sub{s}}

\makeatletter
\@ifundefined{textcolor}{}
{%
 \definecolor{BLACK}{gray}{0}
 \definecolor{WHITE}{gray}{1}
 \definecolor{RED}{rgb}{1,0,0}
 \definecolor{GREEN}{rgb}{0,1,0}
 \definecolor{BLUE}{rgb}{0,0,1}
 \definecolor{CYAN}{cmyk}{1,0,0,0}
 \definecolor{MAGENTA}{cmyk}{0,1,0,0}
 \definecolor{YELLOW}{cmyk}{0,0,1,0}
 }

\makeatother

\usepackage{babel}
\begin{document}

\title{Mott Transition for Strongly-Interacting 1D Bosons in a Shallow Periodic Potential}

\author{G. Bo\'eris}
\affiliation{Laboratoire Charles Fabry, Institut d'Optique, CNRS, Univ Paris Sud
11, 2 avenue Augustin Fresnel, F-91127 Palaiseau cedex, France}

\author{L. Gori}
\affiliation{LENS and Dipartimento di Fisica e Astronomia, Universit\`a di Firenze, and CNR-INO 50019 Sesto Fiorentino Italy}

\author{M. D. Hoogerland}
\affiliation{Department of Physics, University of Auckland Private Bag 92019, Auckland, New Zealand}

\author{A. Kumar}
\affiliation{LENS and Dipartimento di Fisica e Astronomia, Universit\`a di Firenze, and CNR-INO 50019 Sesto Fiorentino Italy}

\author{E. Lucioni}
\affiliation{LENS and Dipartimento di Fisica e Astronomia, Universit\`a di Firenze, and CNR-INO 50019 Sesto Fiorentino Italy}

\author{L. Tanzi}
\affiliation{LENS and Dipartimento di Fisica e Astronomia, Universit\`a di Firenze, and CNR-INO 50019 Sesto Fiorentino Italy}

\author{M. Inguscio}
\affiliation{LENS and Dipartimento di Fisica e Astronomia, Universit\`a di Firenze, and CNR-INO 50019 Sesto Fiorentino Italy}
\affiliation{INRIM, 10135 Torino, Italy}

\author{T. Giamarchi}
\affiliation{Department of Quantum Matter Physics, University of Geneva, 24 quai Ernest-Ansermet, 1211 Geneva, Switzerland}

\author{C. D'Errico}
\affiliation{LENS and Dipartimento di Fisica e Astronomia, Universit\`a di Firenze, and CNR-INO 50019 Sesto Fiorentino Italy}

\author{G. Carleo}
\affiliation{Laboratoire Charles Fabry, Institut d'Optique, CNRS, Univ Paris Sud
11, 2 avenue Augustin Fresnel, F-91127 Palaiseau cedex, France}

\author{G. Modugno}
\affiliation{LENS and Dipartimento di Fisica e Astronomia, Universit\`a di Firenze, and CNR-INO 50019 Sesto Fiorentino Italy}

\author{L. Sanchez-Palencia}
\affiliation{Laboratoire Charles Fabry, Institut d'Optique, CNRS, Univ Paris Sud
11, 2 avenue Augustin Fresnel, F-91127 Palaiseau cedex, France}

\date{\today}

\begin{abstract}
We investigate the superfluid-insulator transition of one-dimensional interacting Bosons in both deep and shallow periodic potentials. We compare a theoretical analysis based on Monte-Carlo simulations in continuum space and Luttinger liquid approach with experiments on ultracold atoms with tunable interactions and optical lattice depth. Experiments and theory are in excellent agreement.
It provides a quantitative determination of the critical parameter for the Mott transition and defines the regime of validity of widely used approximate models, namely the Bose-Hubbard and sine-Gordon models.
\end{abstract}
\maketitle

\begin{bibunit}

\paragraph*{Introduction.---} \label{sec:intro}

The interplay of repulsive interactions and a periodic potential in a quantum fluid triggers a superfluid-insulator transition  known as the Mott transition, provided the potential period is commensurate to the inverse fluid density. The most familiar notion of Mott transition takes place in the limit of a deep periodic potential. In this case the lattice Hubbard model microscopically captures the dominant interaction and hopping processes, the strengths of which, $U$ and $\hopping$, strongly depend on the periodic potential amplitude $V$. Then, the Mott transition is driven by the competition of these sole two parameters at $\hopping \sim U$~\cite{mott1949,mott1990}. Quite strikingly in one dimension (1D) a Mott transition can exist even for a vanishingly small periodic potential provided the repulsive interactions are strong enough~\cite{haldane1980,haldane1981,giamarchi1997,giamarchi2004}.
In the limit of a shallow potential, its amplitude $V$ becomes sub-relevant and the transition is mostly controlled by the interaction
strength $g$ alone~\cite{giamarchi2004}.

Ultracold atoms provide a remarkable laboratory to study this physics~\cite{lewenstein2007,bloch2008}.
So far the Mott transition has been observed in both deep~\cite{greiner2002,stoeferle2003,koehl2005b,jordens2008,schneider2008} and shallow~\cite{haller2010} optical lattices. However, the characterization of the Mott transition in shallow potentials remains a formidable challenge for both theory and experiments,
with direct consequences not only in the ultracold atom realm but also in condensed matter for problems such as spin chains for instance~\cite{oshikawa1997,pouget2001,giamarchi2004}.

On the theoretical side, while the Hubbard limit is now well documented~\cite{lewenstein2007,bloch2008,georges2010} and its Mott transition has been extensively studied~\cite{fisher1989,batrouni1990,jaksch1998,kuhner1998,kuhner2000,capogrosso2007,mark2011}
its regime of validity beyond the deep-lattice limit is still unknown in 1D, with ab-initio results having been reported so-far only for 3 dimensions (3D)~\cite{pilati2013}. In the limit of a vanishing potential, an estimate of the transition values may be found in the sine-Gordon model whose coefficients are determined perturbatively~\cite{buchler2003,haller2010}.
This, however, ignores the unavoidable renormalization of the field-theoretic coupling parameters by the potential, which may significantly affect the transition. On the experimental side, the Mott transition has been clearly observed in the shallow lattice limit using modulation spectroscopy and transport measurements~\cite{haller2010}. However, the experimental uncertainties did not allow to a precise determination of the phase diagram.

In this Letter, we report the first quantitative joint theoretical and experimental investigation of the Mott transition for strongly-interacting 1D Bosons in a shallow periodic potential. Using continuous-space quantum Monte Carlo calculations, we determine the exact quantum phase diagram. Our calculations confirm the field-theoretical universal predictions and provide, in addition, an accurate quantitative values of the critical parameters of the Mott transition. Experimentally, we perform transport measurements on a Bose gas with tunable interactions down to the limit of very shallow lattices and we analyze them with a phase slip based model to determine accurately the Mott transition. The numerical and experimental results are in excellent agreement and show significant deviation from the perturbative sine-Gordon theory using \emph{bare} Luttinger parameters.

\begin{figure*}[t!]
 \centering
 \includegraphics[width=\columnwidth]{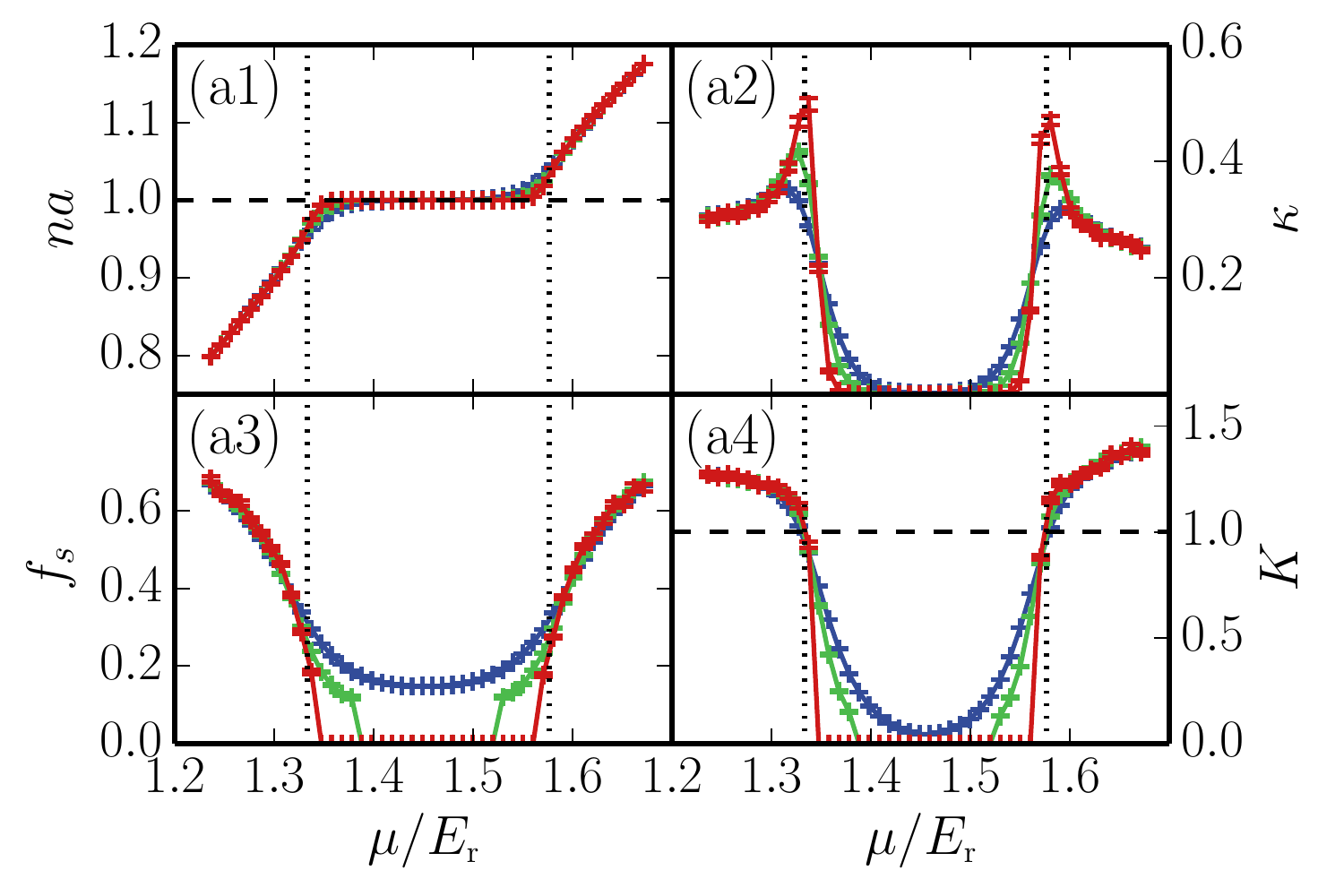}
 \includegraphics[width=\columnwidth]{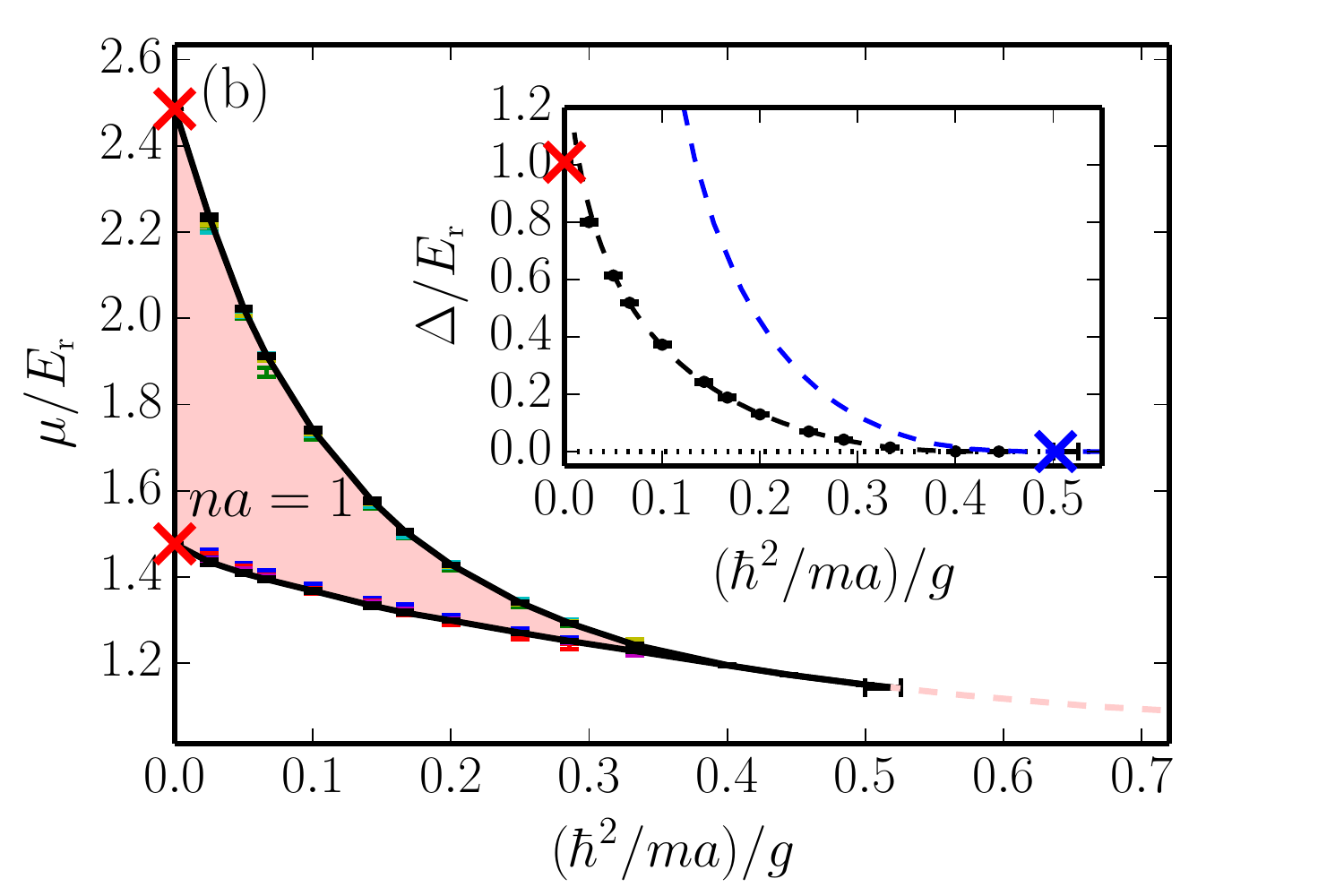}
 \caption{(Color online) (a)
 QMC results for
 the number of particles per site, $na$,
 the compressibility $\kappa$,
 the superfluid fraction $\fs$,
 and the Luttinger parameter $K$,
 versus the chemical potential $\mu$, for a periodic potential amplitude $V = 2\Er$ and an interaction strength $ g = 7\hbar^2/ma$.
 The various curves correspond to increasing system sizes, $L / a = $ 30, 50, and 100 (blue, green, and red respectively).
 The vertical dotted lines show the transition points determined from the criterion $\Kc = 1$.
 (b) Phase diagram in the $g$-$\mu$ plane for the fixed amplitude of the periodic potential $V = 2\Er$.
 The black points joint by lines are determined from the $\Kc = 1$ criterion,
 while the colored points show the results found from
 the crossing point of the compressibility (red),
 the cusp of the compressibility (green),
 and the crossing point of the Luttinger parameter (blue).
 The red crosses are the hard-core limits.
 The inside of the lobe (red region) and the dashed line
 correspond a density of one particle per potential spacing.
 Inset: Mott gap versus interaction strength. The black dashed line is a fit to the BKT prediction (see text).
 The blue dashed line is the prediction of the BH model and the blue cross is the corresponding tip of the lobe.}
 \label{fig:panel}
\end{figure*}

\paragraph*{Model and theoretical approach.---} \label{sec:model}
We consider zero-temperature interacting 1D Bosons of mass $m$ with a contact interaction of strength $g$, subjected to a periodic potential $V(x) = V \sin^2(k x)$ of spacing $a=\pi/k$ and amplitude $V$.
Both the large $V$ and small $V$ limits have the possibility of a Mott transition when the interactions are increased~\cite{georges2010}.
In spite of their qualitative different natures the two limiting cases are, however, expected to belong to the same universality class for they both lead to the same low-energy sine-Gordon model~\cite{haldane1981,giamarchi2004,georges2010,cazalilla2011}. Within the Tomonaga-Luttinger liquid (TLL) approach, the homogeneous superfluid is parameterized by the Luttinger parameter $K$, which characterizes the interaction strength. For weak interactions, the periodic potential is essentially irrelevant, except in renormalizing the effective value of the Luttinger parameter. For strong interactions, the TLL may be unstable upon introducing a periodic potential, which signals the Mott insulator phase. More precisely, the Mott transition may be triggered either by changing the fluid density to commensurability at sufficiently strong interactions (Mott-$\delta$ transition) or by increasing the interactions at commensurability (Mott-$U$ transition). The TLL theory predicts the universal critical values $K_\sub{c}=1/p^2$ and  $K_\sub{c}=2/p^2$ for the Mott-$\delta$ and Mott-$U$ transitions respectively, where $p$ is the commensurability order~\cite{giamarchi1997,giamarchi2004,cazalilla2011}.

However the TLL theory involves effective parameters that are not easily related to the
microscopic Hamiltonian parameters and the critical curve $g_\sub{c}(V)$ is presently
not quantitatively known. To precisely determine the Mott transitions, we use quantum Monte-Carlo (QMC) simulations.
This allows us
(i)~to determine quantitatively the phase diagram in terms of the microscopic parameters
and (ii)~to compute explicitly the Luttinger parameter $K$ as a function of the microscopic ones and make the link with field theory.
We use the same implementation of the continuous-space worm algorithm~\cite{boninsegni2006,boninsegni2006b} in the grand-canonical ensemble as used in Ref.~\cite{carleo2013}, which is numerically exact for all the physical quantities we study in the following~\cite{noteSupplMat}.

\paragraph*{Incommensurate transition.---} \label{sec:incommensurate}
We start with the incommensurate (Mott-$\delta$) transition, which may be triggered by changing the chemical potential $\mu$.
In order to accurately determine the critical point, several quantities are examined.
The particle density $n$ is computed directly in QMC and the compressibility $\kappa \equiv \partial n / \partial \mu$
is computed independently from particle number fluctuations.
The hydrodynamic superfluid density $\ns$ is found from the superfluid stiffness $\Upsilon_\sub{s}$, defined as the response of the system to a twist of the boundary conditions, which is computed using the winding-number estimator~\cite{noteSupplMat}.
We then deduce the superfluid fraction $\fs = \ns / n$ and the Luttinger parameter $K = \pi \sqrt{\ns \kappa}$.

The QMC results are shown versus the chemical potential in \reffig{fig:panel} for $V = 2\Er$
where $\Er = \hbar^2 k^2 / 2m$ is the recoil energy, $g = 7 \hbar^2 / m a$, and various system sizes.
The density [\reffig{fig:panel}(a1)] increases monotonically with the chemical potential $\mu$ and exhibits a plateau at commensurability, $na = 1$, where the superfluid density drops to zero. This is the signature of the Mott-$\delta$ transition.
The critical chemical potentials $\mu^\pm_\sub{c}$ corresponding to the two edges of the plateau are accurately determined from the crossing points of the compressibility for different system sizes [\reffig{fig:panel}(a2)].
They can also be found from the drop of the superfluid fraction $\fs$,
which yields similar values for $\mu_\sub{c}^\pm$ [\reffig{fig:panel}(a3)].
At the Mott-$\delta$ transition, the Luttinger parameter is expected to exhibit the universal discontinuity from $K=1$ to $K=0$. Our data [\reffig{fig:panel}(a4)] are perfectly compatible with this prediction of the TLL theory.

Repeating the same calculations for various values of the interaction strength, we find the Mott lobe in the $g-\mu$ plane shown in \reffig{fig:panel}(b).
The black points and joining lines are determined from the $\Kc = 1$ criterion while the colored points are extracted from
the crossing point of the compressibility (red), the cusp of the compressibility (green),
and the crossing point of the Luttinger parameter (blue).
The different methods yield results in excellent agreement all along the lobe within a few percents.
The Mott gap $\Delta = \mu_\sub{c}^+ - \mu_\sub{c}^- $ is shown in the Inset of \reffig{fig:panel}
versus the interaction strength.
The black dashed line is a fit with the exponential closing $\Delta(g) \propto \exp (-b / \sqrt{g - g_\sub{c}})$ predicted by the Berezinskii-Kosterlitz-Thouless (BKT) theory \cite{berezinskii1971,kosterlitz1973,kashurnikov1996b}, with $b$ and $g_\sub{c}$ as fitting parameters.
We find that the fit is good for all values of $g$ except very close to the hard-core limit, $g \to +\infty$.
The fit also yields a fair estimate of the critical value at commensurability,
$g_\sub{c} / (\hbar^2 / m a) = 2.2 \pm 0.1$, which corresponds to the tip of the lobe.
\begin{figure}[t!]
\centering
\includegraphics[width=\columnwidth]{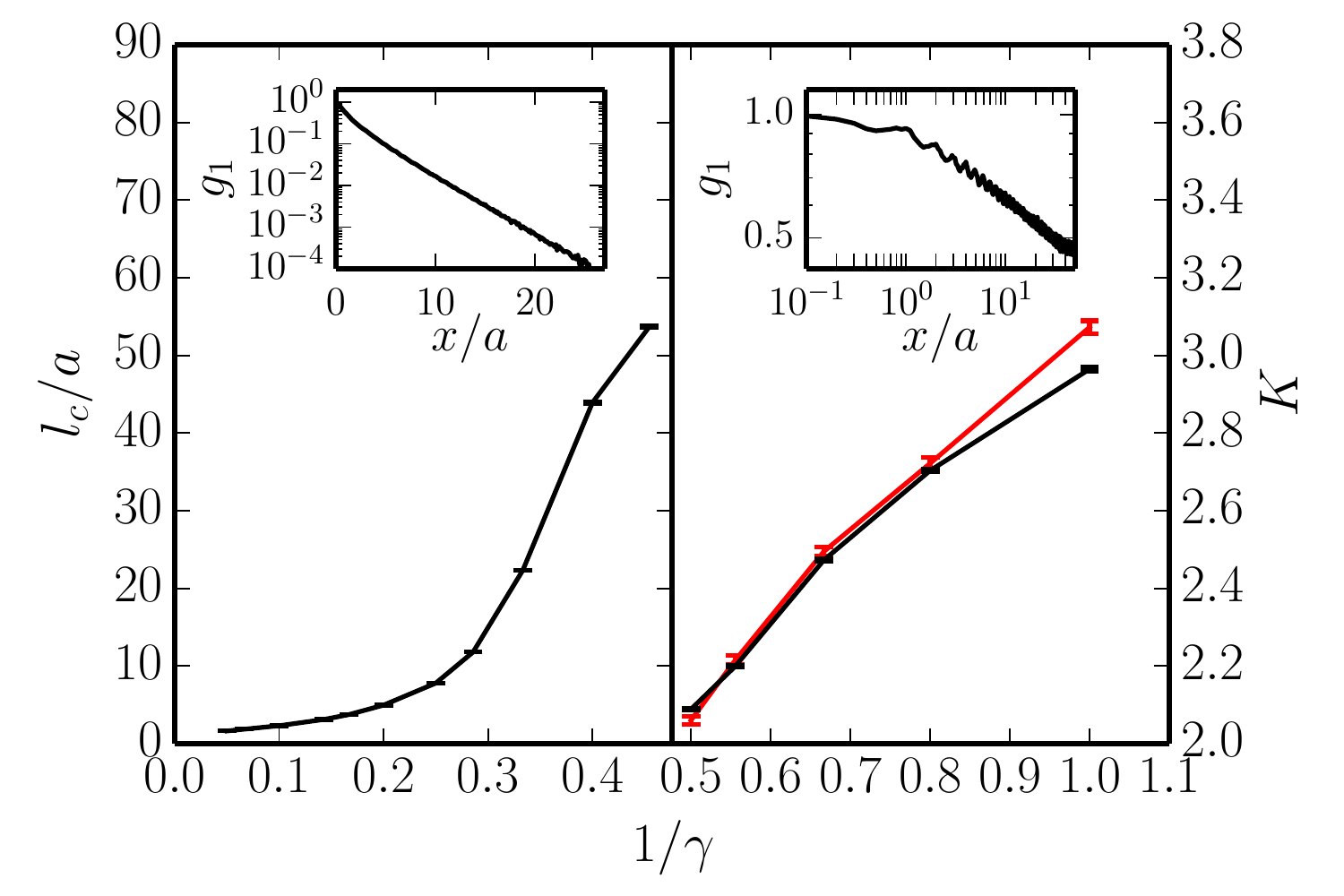}
\caption{(Color online) Analysis of the correlation function
$g_1(x) = \braket{ \hat \psi^\dagger (x) \hat \psi(0) }$
for $V / \Er = 2$. The decay is exponential in the insulator (left inset, $\gamma = 7$)
  and algebraic in the superfluid (right inset, $\gamma = 1.25$).
  Left: Correlation length in the insulator.
  Right: Luttinger parameter in the superfluid extracted from the decay of
  the correlation function (black) and from the formula $K = \pi \sqrt{\ns \kappa}$ (red).
  The QMC results are the points, connected by straight lines to guide the eye.
  Note that the inflection of the correlation length curve close to the transition is due to finite size effects.
  }
  \label{fig:correlations}
  \end{figure}

\paragraph*{Commensurate transition.---} \label{sec:commensurate}
In order to characterize the commensurate (Mott-$U$) transition more precisely, we now vary the interaction strength $\gamma \equiv mg/\hbar^2n$ at commensurability, $na = 1$.
We compute the one-body correlation function
$g_1(x) = \langle \hat \psi^\dagger (x) \hat \psi(0) \rangle$,
where $\psi$ is the field operator,
from the statistics of endpoints on open world lines in the QMC~\cite{boninsegni2006,boninsegni2006b}.
When increasing $\gamma$ along the line with $na=1$ (dashed line in \reffig{fig:panel}),
we observe a clear change of behavior of the $g_1$ function
from algebraic for $\gamma<\gamma_\sub{c}$ to exponential for $\gamma>\gamma_\sub{c}$ (see Insets of \reffig{fig:correlations}). This is the signature of the Mott-$U$ transition.
The finite correlation length $l_\sub{c}$ in the insulating phase in shown in the left panel of \reffig{fig:correlations}. It is of only a few lattice sites long for strong interactions and increases up to a value comparable to the system size for $\gamma_\sub{c} \sim 2$.
This is compatible with the expected divergence of the correlation length at the transition.
In the superfluid phase, the algebraic decay of the correlation function is compatible with the TLL theory prediction $g_1(x) \propto 1 / x ^ {1 / 2K}$.
The two values of the Luttinger parameter found from a fit to this prediction and from the thermodynamic prediction $K=\pi\sqrt{\ns\kappa}$ are in good agreement (see right panel of \reffig{fig:correlations}).
When increasing the interaction towards the insulating phase,
the Luttinger parameter decreases down to $K \simeq 2$
as predicted by the TLL theory.

\begin{figure}[t!]
 \centering
 \includegraphics[width=\columnwidth]{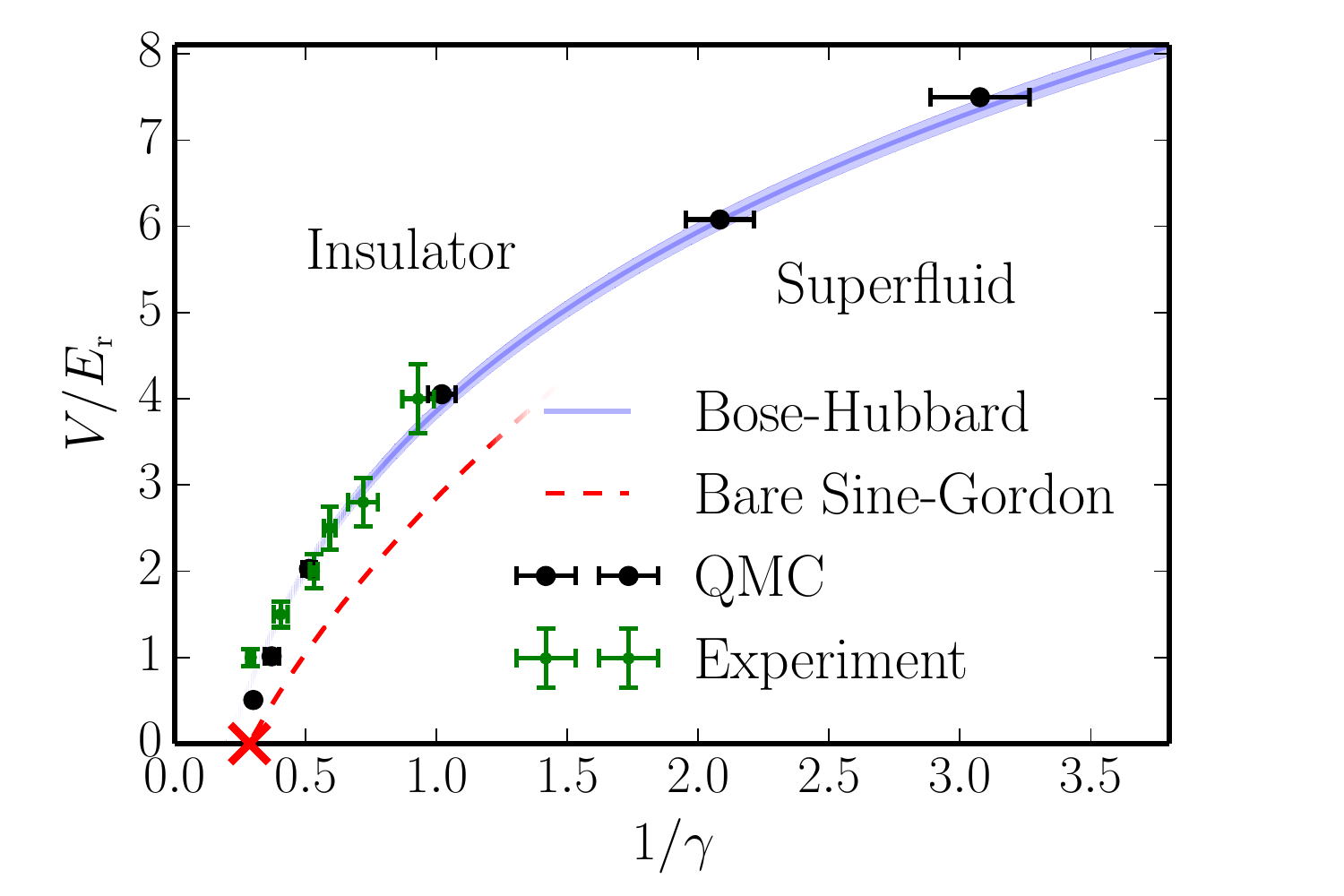}
 \caption{(Color online) Phase diagram in the $g$-$V$ plane at commensurability ($na=1$).
 The black and green points, both with errorbars) are the Monte Carlo and experimental results, respectively.
 The blue line is the BH prediction using exact Wannier functions and the critical value $(J / U)_\sub{c} = 0.297 \pm 0.01$ (the shaded area corresponds to this error bar).
 The red cross indicates the critical interaction strength $g_\sub{c}$ such that $K(g_\sub{c}) = 2$ in the Lieb-Liniger model.
Along the whole transition line the \emph{effective} Luttinger parameter remains $K=2$.
The red line is the result of the \emph{bare} sine-Gordon theory~\cite{buchler2003}.}
 \label{fig:phase_diagram}
\end{figure}

To locate the Mott-$U$ transition point accurately, we resort to the BKT renormalization group
equations to perform the finite size scaling of the Luttinger parameter~\cite{noteSupplMat}.
The results are shown in \reffig{fig:phase_diagram} (black points).
In the strong potential limit, the results are compatible with the prediction of the BH model
with the critical value $(\hopping / U)_\sub{c} = 0.297 \pm 0.01$~\cite{kuhner1998,kuhner2000} and the hopping $\hopping$ and interaction strength $U$ calculated from the exact Wannier functions~\cite{jaksch1998}.
In the vanishing potential limit, the results converge to the critical value $\gamma_\sub{c}(V=0) \simeq 3.5$ (red cross in \reffig{fig:phase_diagram}) found from the exact relation $K(g)$ for the integrable Lieb-Liniger model~\cite{lieb1963a}.
However, in the intermediate regime, we find a strong deviation from the pinning transition line
(red dashed line) computed in Ref.~\cite{buchler2003} from the perturbative sine-Gordon theory using the \emph{bare} Luttinger parameters. This shows that to quantitatively obtain the phase diagram the renormalization of the Luttinger parameters by even relatively weak interactions is significant and cannot be ignored in the perturbative field theory.

\paragraph*{Experiment.---}
Having quantitatively characterized the Mott transition from theory, we now turn to the experimental investigation. The experiment starts with a Bose-Einstein condensate of 35\,000 $^{39}$K atoms with tunable scattering length at a broad Feshbach resonance~\cite{roati2007}. The BEC is  split into about 1000 vertical 1D tubes by adiabatically loading  a strong horizontal 2D optical lattice. Each tube contains on average 36 atoms and the transverse trapping frequency, $\omega_{\perp}=2 \pi \times 40$ kHz corresponds to an energy higher than all other energy scales, realizing an effective 1D geometry. In the longitudinal direction we then adiabatically rise a weak optical lattice with spacing $a=\lambda/2=$532 nm and normalized amplitude $V/\Er$ ranging from 1.0(1) to 4.0(4). The temperature of the system, $T\simeq30$\,nK, is below the 1D degeneracy temperature $T_\sub{D}\simeq50$\,nK~\cite{petrov2000b}. A magnetic field along the vertical direction holds the system against gravity and a residual harmonic trap potential, with frequency $\omega_z=2 \pi \times 160$ Hz, makes it inhomogeneous. By varying the 3D scattering length $a_\sub{3D}$, we can tune the Lieb-Liniger parameter $\gamma$ in the range $0.07-7.4$. The system parameters are chosen to obtain a mean atom number per potential period $\langle na\rangle=$1. This implies that in most of the tubes there are one or two regions with local commensurate density, $na$=1, which can undergo a Mott-$U$ transition. There is however a fraction of the tubes for which $na<1$ which cannot become insulating~\cite{noteSupplMat}.

\begin{figure}[t!]
\centering
{\includegraphics[width=0.95\columnwidth] {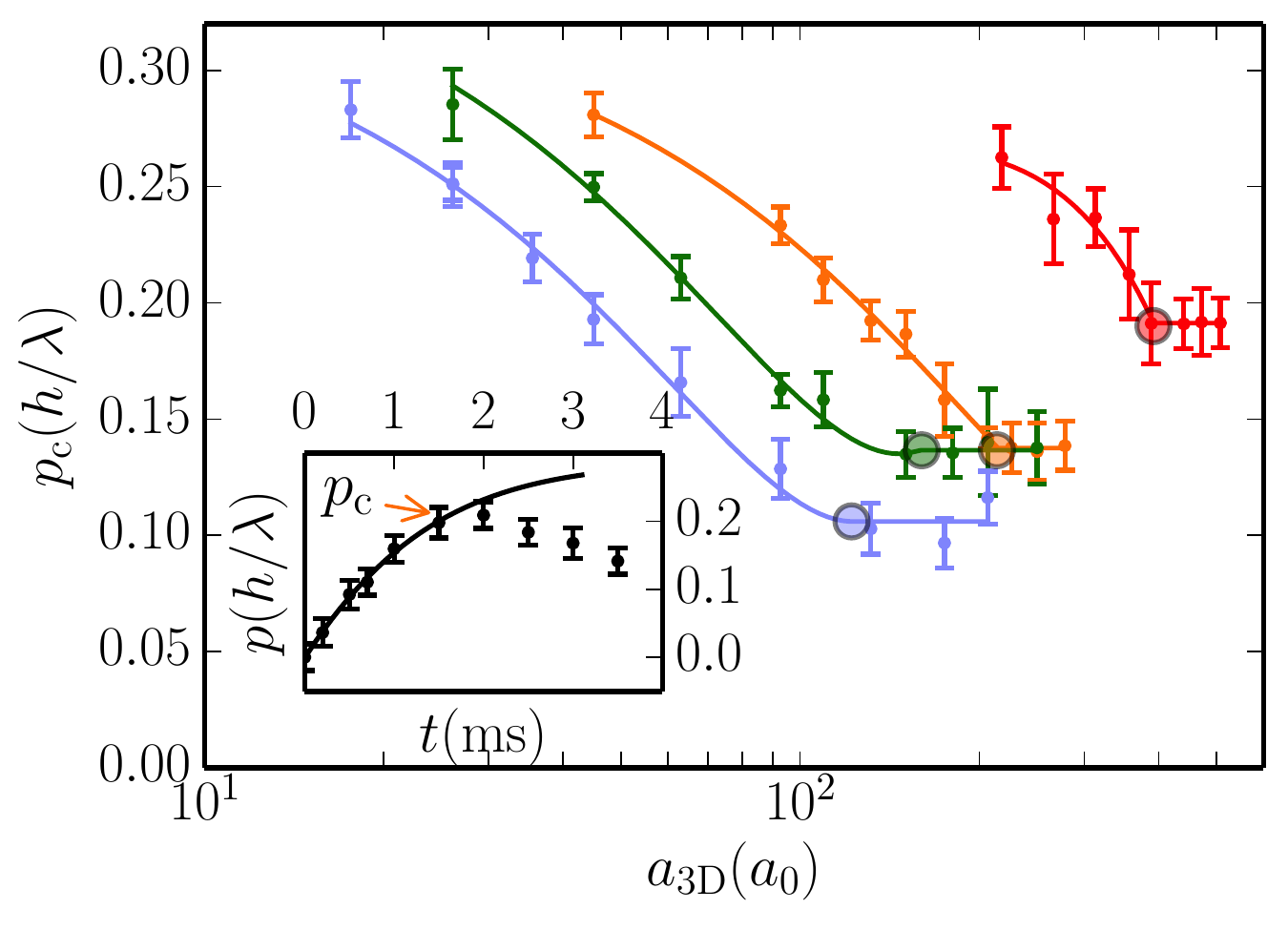}}
\caption{(Color online) Critical momentum $p_\sub{c}$ as a function of the scattering length $a_\sub{3D}$ for four periodic potential depth values: $V/\Er$=1 (red), 2 (orange), 2.8 (green), 4 (blue). A piecewise second-order polynomial fit determines the critical values  for the SF-MI transition: respectively $a_\sub{c}/a_0$=$392\pm 12$, $214\pm 6$, $160\pm 30$, $122 \pm 8$. Inset: time evolution of the momentum distribution peak $p$ for $a_\sub{3D}=$109$\,a_0$ and $V/\Er$=2. The solid line is the theoretical damped oscillation fitting the data $p<p_\sub{c}$ before the dynamical instability sets in. The error bars represent the root mean square of the imaging resolution and statistical uncertainties.}
\label{fig_Exp}
\end{figure}

To detect the Mott transition we excite a sloshing motion of the system through a shift of the trapping potential, obtained by suddenly switching off the magnetic field gradient~\cite{haller2010,tanzi2013}. We let the atoms evolve in the trap for a variable time $t$, after which all optical potentials are switched off and time-of-flight absorption images are recorded. An example of the time evolution of the momentum distribution peak $p(t)$ is shown in the inset of~\reffig{fig_Exp}: One can notice an initial increase of $p$ followed by a subsequent decrease. We analyze this behavior in the frame of a phase slip based model~\cite{polkovnikov2005,tanzi2013}. Phase slips, i.e.\;the dominant  excitations in 1D, make the system dynamics  dissipative already at small momenta: $p(t)$  can be fit with a damped oscillation function of the form $p(t)=p_\sub{max} e^{-G t}\sin(\omega't)$, where the frequency $\omega'$ is renormalized by the damping rate $G$ and by the lattice \cite{noteSupplMat}. At larger momenta one observes a clear deviation from such behavior, with a sudden increase of the damping that significantly reduces the growth of $p$. We identify the momentum  where the experimental data points deviate with respect to the theoretical curve as the critical momentum $p_\sub{c}$ for the occurrence of a dynamical instability, driven by a divergence of the phase slip rate~\cite{tanzi2013}. The critical momentum $p_\sub{c}$ is expected to vanish at the superfluid-insulator transition~\cite{polkovnikov2005}.

The behavior of $p_\sub{c}$ as a function of the scattering length $a_\sub{3D}$ is reported on \reffig{fig_Exp} for several values of the lattice depth. The measured $p_\sub{c}$ initially decreases for increasing $a_\sub{3D}$, and then reaches a finite constant value. We interpret the onset of the plateau as the Mott transition for the commensurate regions of the system, which drives all the corresponding tubes into an effective insulating regime (i.e. transport along individual tubes is globally suppressed). The fraction of tubes that does not reach the critical density $na$=1 keeps instead moving also beyond this point, originating the observed plateau for $p_\sub{c}$.  This interpretation is corroborated by the observed increase of $p_\sub{c}$ at the plateau for decreasing $V$, since the increase of the interaction strength that is necessary to reach the insulating regime produces an overall decrease of the density of the 1D systems, hence an increase of the fraction of tubes that does not reach $na$=1. For each set of measurements with a given value of $V$, we therefore identify the critical scattering length value $a_\sub{c}$ to enter the insulator regime by determining the beginning of the plateau with a piecewise fit. We use a second-order polynomial fit, which is justified by the phase slip based model~\cite{polkovnikov2005,danshita2012,noteSupplMat}. We clearly see that as $V$ decreases,  $a_\sub{c}$ -- and thus also $\gamma_\sub{c}$ -- increases.

For each periodic potential depth we get the Mott-$U$ transition point converting $a_\sub{c}$ into $\gamma_\sub{c}$ for $na=1$. The experimental results are shown as green points in \reffig{fig:phase_diagram}. Within our uncertainties~\cite{noteExp}, the experiment is in very good agreement with the numerical simulations, except for the case $V/\Er$=1, where the finite size of the system might start to play a role. Our results are also consistent with those reported in Ref.~\cite{haller2010} within their uncertainties.
Our experiment confirms the clear deviation of the transition line from the bare sine-Gordon prediction observed in the QMC results. Note that, surprisingly enough, while the BH model is justified only for $V \gg \Er$, both numerics and experiments show that the BH prediction for the Mott-$U$ transition is quite accurate down to the limit $V \rightarrow 0$. This agreement is however rather accidental and the breakdown of the BH model is manifest in other quantities. For instance, the BH prediction for the Mott gap deviates significantly from the exact QMC result [see Inset of \reffig{fig:panel}(b)].

\paragraph*{Conclusions.---}
We have studied, both theoretically and experimentally, the Mott transition of strongly-interacting 1D Bosons in a periodic potential from deep to shallow potentials. Our ab-initio QMC calculations validate the field-theoretic universal predictions and in addition provide a quantitative determination of the phase diagram.
It shows that the renormalization of the Luttinger parameter is significant even for weak periodic potentials. The numerical analysis give excellent agreement with experiments for the Mott-$U$ transition.
The experimental observation of our numerical phase diagram for the Mott-$\delta$ transition is still beyond reach for ultracold atomic systems due to the requirement of a fine control of atom number in box-shaped potentials. In spite of recent progress in that direction~\cite{meyrath2005,gaunt2013}, it remains a great challenge for future studies.

\begin{acknowledgments}
This research was supported by
the EU FET-Proactive QUIC (H2020 grant No.~641122),
the ERC ALoGlaDis (FP7/2007-2013 grant No.~256294),
Marie Curie IEF (FP7/2007-2013 grant No.~327143),
MIUR (grant No.~RBFR12NLNA),
and the Swiss NSF under Division II.
Numerical calculations were performed using HPC resources from GENCI-CCRT/CINES (grant No.~c2015056853)
and the GMPCS cluster of the LUMAT federation (FR LUMAT 2764),
and make use of the ALPS scheduler library and statistical analysis tools~\cite{troyer1998,ALPS2007,ALPS2011}.

\paragraph*{Note added.---}
During the completion of this manuscript, a preprint appeared reporting the numerical study of the Mott-$U$ transition with results consistent with ours~\cite{astrakharchik2015}.
\end{acknowledgments}

\bibliographystyle{apsrev}

\end{bibunit}

\begin{bibunit}

\renewcommand{\theequation}{S\arabic{equation}}
\setcounter{equation}{0}
\renewcommand{\thefigure}{S\arabic{figure}}
\setcounter{figure}{0}
\onecolumngrid  
    
\pagebreak

\newpage

{\center \bf \large --\emph{Supplemental Material}--\\
Mott Transition for Strongly-Interacting 1D Bosons in a Shallow Periodic Potential\\}

\vspace*{0.5cm}

In this Supplemental Material we give additional details on the methods and the analysis we
have performed in the main Letter. In particular we provide a detailed description of
i)~the quantum Monte Carlo calculations,
ii)~the determination of the superfluid density,
iii)~the finite-size scaling analysis, and
iv)~the experimental methods.

\vspace*{0.5cm}

\twocolumngrid  

\subsection*{Quantum Monte Carlo calculations} \label{sec:qmc_methods}
We consider 1D interacting Bosons at zero temperature governed by the continous-space Hamiltonian,
\begin{equation}
\hat H = \int \!\! \dd x \, \left[ \frac{\hbar^2}{2m} \nabla \hat \psi^\dagger \nabla \hat \psi
+  \frac{g}{2} \hat \psi^\dagger \hat \psi^\dagger \hat \psi \hat \psi + V(x) \hat \psi^\dagger \hat \psi \right],
\label{eq:hamiltonian}
\end{equation}
where $\hat\psi(x)$ is the Bose field,
$m$ is the particle mass,
$g$ is the contact interaction strength,
and $V(x) = V \sin^2(k x)$ is a periodic potential of spacing $a=\pi/k$ and amplitude $V$.

We study the interacting quantum system from first principles, by means of quantum Monte Carlo (QMC) numerical simulations. 
Path-integral QMC approaches allow us to compute the exact thermodynamic properties of a generic, interacting bosonic quantum system in any dimensions.
This is achieved by means of the path-integral representation of the grand-canonical partition function $Z = \Tr [e^{-\beta (\hat H - \mu \hat N)}]$, where $\beta = 1/\kB T$ with $\kB$ the Boltzman constant and $T$ the temperature, $\mu$ is the chemical potential, and $\hat N = \int \!\! \dd x \, \hat \psi^\dagger \psi$ is the total particle number operator. 
The quantum partition function is expressed as an equivalent classical partition function of interacting polymers living in dimension $D+1$, where the additional dimension is the imaginary-time direction~\cite{ceperley1995}. The equivalent classical partition function can be treated stochastically by means of Monte Carlo sampling. An efficient way of sampling the associated partition function is given by the worm algorithm~\cite{boninsegni2006,boninsegni2006b}.

In our implementation, the imaginary-time propagator entering the path-integral representation is written as
\begin{multline}
 \braket{ \{ \vec r_i \} | e^{-\epsilon \hat H} | \{ \vec r'_i \} } =
            \prod_{i = 1}^N \rho^{(0)}_\epsilon (\vec r_i, \vec r'_i)
            \prod_{i = 1}^N e^{-U^{(1)}_\epsilon (\vec r_i, \vec r'_i)} \times \\
            \times \prod_{i < j} e^{-U^{(2)}_\epsilon (\vec r_i, \vec r_j, \vec r'_i, \vec r'_j)},
\end{multline}
where $\rho^{(0)}_\epsilon (\vec r_i, \vec r'_i) = \sqrt{m / 2 \pi \hbar^2 \epsilon} \times e^{-m |\vec r_i - \vec r'_i|^2 / 2 \hbar^2 \epsilon}$
is the free-particle propagator. The one-body contribution
$e^{-U^{(1)}_\epsilon (\vec r_i, \vec r'_i)} =
   \braket{\vec r_i | e^{-\epsilon [\hat{ \vec p}^2 / 2m + V(\hat{\vec r})]} | \vec r'_i} / \rho^{(0)}_\epsilon(\vec r_i, \vec r'_i) $ is computed
exactly by matrix squaring~\cite{ceperley1995}. 
The two-body interaction is taken into account at the pair-product level, thanks to the explicit expression for the two-body propagator 
$e^{-U^{(2)}_\epsilon (\vec r_i, \vec r_j, \vec r'_i, \vec r'_j)} = 
   \braket{ \vec r_i, \vec r_j | e^{-\epsilon[ (\hat{ \vec p}_i^2 + \hat{ \vec p}_j^2) / 2m ) + g \delta(\hat{\vec r}_i - \hat{\vec r}_j)]} | \vec r'_i, \vec r'_j}
    / \rho^{(0)}_\epsilon (\vec r_i, \vec r'_i) \rho^{(0)}_\epsilon (\vec r_j, \vec r'_j) $
which is known~\cite{gaveau1986}. 
The systematic error coming from the discretization of the path-integral along the imaginary-time direction is smaller than the statistical errorbars reported in our results.

\subsection*{Superfluid density} \label{sec:superfluid_density}
The stiffness $\Upsilon_\sub{s} = -\left.\frac{mL}{\beta \hbar^2} \frac{\partial^2 \ln Z}{\partial \theta_0^2} \right|_0$
measures the response of the system when the periodic boundary conditions are twisted by an angle $\theta_0$.
It is directly computed by QMC using the usual winding number estimator~\cite{ceperley1995}.
However, this quantity strongly depends on the ratio $\beta / L$~\cite{prokofev2000}. The relevant quantity to study is the 
superfluid density $\ns$ appearing in the hydrodynamic action
\begin{equation}
 S[\theta] = \int_0^L \dd x \int_0^{\hbar \beta} \dd \tau \left[ \frac{\hbar^2 \ns}{2m} (\partial_x \theta)^2 
 + \frac{\hbar^2 \kappa}{2} (\partial_\tau \theta)^2 \right].
\end{equation}
Computing the partition function $Z = \int \mathcal{D} \theta \, e^{-S[\theta] / \hbar}$, correctly taking into account
winding configurations of the form $\theta(x, \tau) = 2 \pi k x / L$ with $k \in \mathbb{Z}$, one deduces the relation
\begin{equation}
 \Upsilon_\sub{s} = \ns \left(1 - 4 \pi^2 \ns \frac{\hbar^2 \beta}{m L} \frac{\sum_{k = -\infty}^{+\infty} k^2 e^{-2 \pi^2 \ns k^2 \hbar^2 \beta / m L} }{ \sum_{k = -\infty}^{+\infty} e^{-2 \pi^2 \ns k^2 \hbar^2 \beta / m L} } \right).
\end{equation}
In our calculations, we invert this relation and extract $\ns$ from the stiffness $\Upsilon_\sub{s}$ computed by QMC.

\subsection*{Finite size scaling} \label{sec:superfluid_density}
To precisely determine the Mott-$U$ critical interaction strength $g_\sub{c}$, we study the evolution of the Luttinger parameter $K$
for the increasing system sizes $L/a = 30$, $50$, and $100$. When $L$ is increased, the temperature $T$ is lowered by keeping the ratio $\beta / L$ constant in order to consistently compute ground-state properties. 
We resort to the known Berezinkii-Kosterlitz-Thouless (BKT) renormalization group (RG) equations to perform the finite size scaling and determine the critical interaction strength in the thermodynamic limit.
In particular, the finite-size behaviour of the Luttinger parameter $K$ and of renormalized potential strength $V$ are governed by the renormalization equations~\cite{berezinskii1972,kosterlitz1973,prokofev2001,prokofev2002,giamarchi2004},
\begin{equation}
 \frac{\dd K}{\dd \ell} = -\frac{\pi^6}{16} \left( \frac{V}{\Er} \right)^2 K^2
 \quad
 ;
 \quad
 \frac{\dd V}{\dd \ell} = (2 - K) V.
\end{equation}
Dividing these two equations one by another, one finds that the quantity
\begin{equation}
\zeta \equiv \frac{2}{K} + \ln \left( \frac{K}{2} \right) - \frac{\pi^2}{16} \left( \frac{V}{\Er} \right)^2
\end{equation}
is conserved along the RG trajectories.
The separatrix between the superfluid and the insulating
phases then corresponds to $K=2$ and $V=0$, \ie\ $\zeta = 1$. 
In practice, we use the values the Luttinger parameter $K_1$ and $K_2$ for two system sizes $L_1$ and $L_2$, and compute
$\zeta$ using the integral relation
\begin{equation}
 \int_{K_2}^{K_1} \frac{\dd K}{K^2 (\ln K/2 - \zeta) + 2K} = \ln({L_2}/{L_1}).
\end{equation}
The Mott-$U$ critical point is then determined using the criterion $\zeta = 1$.

\subsection*{Experimental methods}
In the experiment, a Bose-Einstein condensate  of $^{39}$K atoms is split in about $10^3$ 1D subsystems (potential tubes) by means of a 2D optical lattice.  The overall Thomas-Fermi distribution of atoms in the 
tube labelled by $i$ and $j$ along the two horizontal directions is given by $N_{i,j}=N_{0,0} [1-2\pi N_{0,0} (i^2+j^2)/5N_\sub{T} ]^{3/2}$, where $N_\sub{T}=3.5 \times10^4$ is the total atom number and  $N_{0,0}\simeq50$ is the atom number in the central tube. On $N_\sub{T}$ there is a 6\% statistical error while a 30\% systematic uncertainty is due to the imaging calibration.
Along the vertical direction of the tube, where we load the weak periodic potential, the trapping frequency is $\omega_z=2 \pi \times 160$ Hz.
A good estimate of the mean atomic density for each tube is provided by the largest of the Thomas-Fermi and the Tonks value~\cite{dunjko2001}. 
The mean site occupation $\langle n a \rangle$ is then calculated by averaging over all the tubes.
During the lattice loading we employ an optimal value of the 3D scattering length to get commensurability on average, \ie\ $\langle n a \rangle=1$. Given $\omega_z$ and $N_\sub{T}$, the optimal value is  $a_{3\textrm {D}}=220\,a_0$.

After the lattice loading, by varying the 3D scattering length $a_{3\textrm{D}}$, we tune the 1D scattering length $a_{1\textrm{D}}=\ell^{2}_{\perp}(1 - 1.03a_{3\textrm{D}}/\ell_{\perp})/2a_{3\textrm{D}}$, and thus the Lieb-Liniger parameter $\gamma$, where $\ell_{\perp}=\sqrt{\hbar /m\omega_{\perp}}$ is the harmonic transverse size of a tube, fixed by the 2D lattice. The value of $\gamma$ is estimated by averaging over all the tubes $1/(n_{max}a a_{1D})$, where the peak occupation $n_{max}a$ is the largest of the Thomas-Fermi and the Tonks value.    
The mean peak occupation, \ie\ the peak occupation averaged over all the tubes, is $\bar{n}_\sub{max}a\simeq1.2$ and the peak density in the central tube is less than two, thus preventing localization mechanisms at occupations other than one. Despite the inhomogeneity of our system, at sufficiently strong interactions we clearly observe a suppression of the system dynamics. We interpret the latter as due to the fact that within each tube, a part of the atoms reaches the localization condition $na=1$ stopping also the remaining adjacent parts with different occupation. We estimate that about one quarter of the atoms resides in tubes where the occupation is always  $na<1$, justifying the plateau in $p_\sub{c}$ shown in Fig.~4 of the main paper. 

The complete model we use to fit the time dependence of the momentum, $p(t)$, as shown in Fig.~4 of the main paper, is $p(t)=p_\sub{max} e^{-G t}\sin(\omega't)$ with amplitude $p_\sub{max} = m^*\omega^{*2}z_0/\omega'$, frequency $\omega' = \sqrt{\omega^{*2} -G^{2}}$ and damping rate $G$. Here $m^*$ is the effective mass due to the shallow lattice, $z_0 \simeq 3\,\mu$m is the trap displacement, and  $\omega^*=\omega_z \sqrt{m/m^*}$ is the lattice renormalized frequency. 

\begin{figure}[htbp]
\includegraphics[width=0.9\columnwidth,clip]{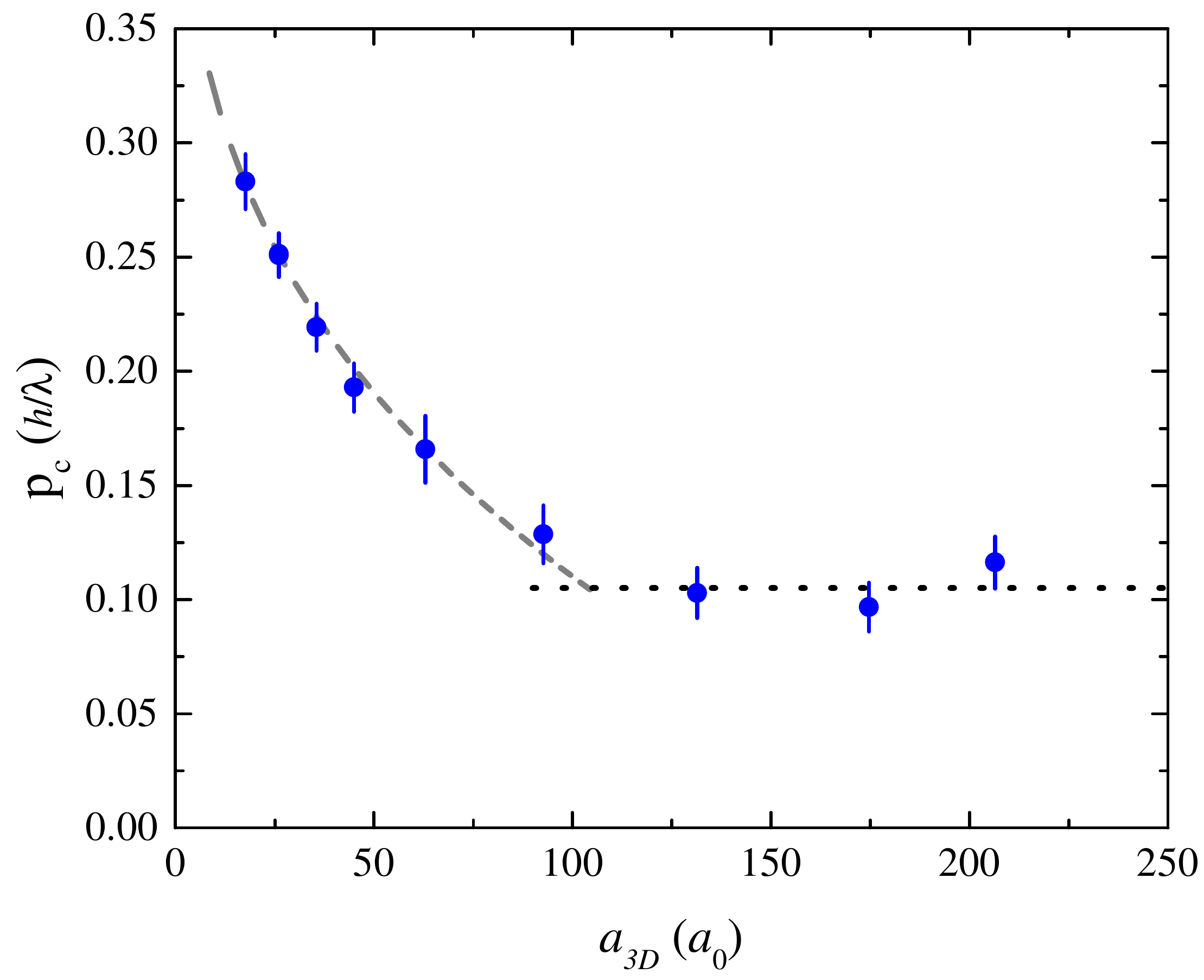}
\caption{For $V/E_\sub{R}=4$ the experimental critical momentum (blue dots) is compared to the one predicted by the quantum phase slip based model, which applies up to the critical scattering length $a_\sub{c}$ to enter the MI regime (gray dashed line). For $a>a_\sub{c}$, $p_\sub{c}$ has instead a  plateau due to tubes with occupation $na<1$ (black dotted line).}.
\label{fig:qps}
\end{figure}

In the absence of an exact theoretical model for the critical momentum $p_\sub{c}$ to enter the dynamical instability regime at finite interaction, we use a quantum phase slips based model to predict the interaction dependence of $p_\sub{c}$. Along the lines of Ref.~\cite{tanzi2013} we use the equation for the quantum phase slip nucleation rate~\cite{polkovnikov2005,danshita2012}
\begin{equation}
\begin{split}
\Gamma=&B\,L(U)\sqrt{naJU}\sqrt{\pi/2-p\lambda/2\hbar}\times\\&\sqrt{\frac{7.1(\pi/2-p\lambda/2\hbar)^{5/2}}
{2\pi\sqrt{U/naJ}}}\times\\&\exp\left[-7.1\sqrt{naJ/U}(\pi/2-p\lambda/2\hbar)^{5/2}\right]\,.
\end{split}
\label{eq.QPS}
\end{equation}
Here $L(U)\simeq 2U^{1/3}$ is the mean length of the tubes and $B$ is a phenomenological constant. Assuming that the system enters  the strongly dissipative regime when the nucleation rate  $\Gamma$ exceeds a constant critical value, from Eq.~(\ref{eq.QPS}) we can obtain $p_\sub{c}$ for each given value of the interaction $U$. Since this relation is valid in the Bose-Hubbard regime, we can employ it to estimate the critical momentum $p_\sub{c}(U)$ for the experimental case with $V/E_\sub{R}=4$. The phenomenological parameter $B$ is arbitrarly adjusted to reproduce the measured $p_\sub{c}$ at one interaction value, $U=2.4J$. In \reffig{fig:qps} we compare the measured and the predicted $p_\sub{c}$ as a function of the scattering length $a_{3\textrm{D}} \propto U$. The quantum phase slip-based model well reproduces the experimental behaviour of $p_\sub{c}$ with the interaction,  showing a quadratic dependence of $p_\sub{c}$ on $a_{3\textrm{D}}$. This result justifies the choice of a quadratic polynomial fit in Fig.~4 of the main paper, at least for the measurements with larger $V$. For shallower lattices the Tomonaga-Luttinger Liquid  model for the nucleation rate could in principle be used~\cite{polkovnikov2005}, but this would require the Luttinger parameter $K$ as a function of the interaction $\gamma$ in the presence of the lattice, which is still unknown.

\bibliographystyle{apsrev}

\begin{thebibliography}{47}
\bibitem[{not({\natexlab{b}})}]{noteExp}
\bibinfo{note}{{The horizontal error bars derive from the fitted $a_\sub{c}$
  while the vertical ones are due to the 10\% of uncertainty in the Raman-Nath
  diffraction technique employed for the calibration of periodic potential
  amplitude}}.
\end{thebibliography}

\end{bibunit}

\end{document}